\documentclass[a4paper,12pt]{article}
\usepackage{amssymb,latexsym}
\usepackage[dvips]{graphicx}

 \DeclareFixedFont{\itshape}{OT1}{cmr}{m}{it}{12}
 
\newtheorem{prp}{Proposition}
\newtheorem{lem}[prp]{Lemma}\newtheorem{thm}[prp]{Theorem}
\newtheorem{cor}[prp]{Corollary}\newtheorem{cnj}[prp]{Conjecture}
\newenvironment{prf}{\begin{trivlist}\item[\emph{Proof.}]}{\end{trivlist}
  \medskip\par}
\newenvironment{prfof}[1]{\begin{trivlist}\item[\emph{Proof of #1.}]}{
  \end{trivlist} \medskip \par}
\newenvironment{rem}{\begin{trivlist}\item[\emph{Remarks.}]}{\end{trivlist}
  \medskip\par}
\def\prpb{\begin{prp}}\def\prpe{\end{prp}}
\def\lemb{\begin{lem}}\def\leme{\end{lem}}
\def\thmb{\begin{thm}}\def\thme{\end{thm}}
\def\corb{\begin{cor}}\def\core{\end{cor}}
\def\cnjb{\begin{cnj}}\def\cnje{\end{cnj}}
\def\prfb{\begin{prf}}\def\prfe{\end{prf}}
\def\prfofb#1{\begin{prfof}{#1}}\def\prfofe{\end{prfof}}
\def\remb{\begin{rem}}\def\reme{\end{rem}}
\def\prpa#1{\label{p:#1}}\def\prpu#1{Proposition~\ref{p:#1}}
\def\lema#1{\label{l:#1}}\def\lemu#1{Lemma~\ref{l:#1}}
\def\thma#1{\label{t:#1}}\def\thmu#1{Theorem~\ref{t:#1}}
\def\cora#1{\label{c:#1}}
\def\seca#1{\label{s:#1}}\def\secu#1{Section~\ref{s:#1}}
\def\eqnb{\begin{equation}}\def\eqne{\end{equation}}
\def\arrb#1{\begin{array}{#1}}\def\arre{\end{array}}
\def\eqna#1{\label{e:#1}}\def\eqnu#1{(\ref{e:#1})}
\def\itmb{\begin{enumerate}}\def\itme{\end{enumerate}}

\def\QED{\relax\ifmmode\let\@tempa\relax\ifcase\@eqcnt\def\@tempa{& & &}\or
  \def\@tempa{& &}\else\def\@tempa{&}\fi\@tempa $\square$ \else\hfill
  $\square$ \fi}
\def\DDD{\relax\ifmmode\let\@tempa\relax\ifcase\@eqcnt\def\@tempa{& & &}\or
 \def\@tempa{& &}\else\def\@tempa{&}\fi\@tempa $\Diamond$
 \else\hfill $\Diamond$ \fi}
\def\dsp{\displaystyle}
\def\limf#1{\displaystyle \lim_{#1\to\infty}}

\def\grad{\mathop{\mathrm{grad}}\nolimits}
\def\diff#1#2{\dsp\frac{d\,#1}{d#2}}
\def\pderiv#1#2{\dsp\frac{\partial\,#1}{\partial#2}}

\def\eps{\epsilon}

\def\interior#1{{#1}^o}
\def\le{\leqq} \def\ge{\geqq} 

\def\reals{{\mathbb R}}
\def\preals{{\mathbb R_+}}

\title{
Uniqueness of fixed point of a two-dimensional map
obtained as a generalization of the renormalization group map
associated to the
self-avoiding paths on gaskets.
}
\author{
Tetsuya Hattori\thanks{
Mathematical Institute, Graduate School of Science, Tohoku University,
 Aoba-ku, Sendai 980--8578, Japan.
E-mai: hattori@math.tohoku.ac.jp
}
}
\date{2006/10/04}
\begin{document}
\maketitle

\begin{center} \textbf{\large ABSTRACT} \end{center}
Let 
$\dsp
W(x,y) = a\,x^3 + b\,x^4 + f_5\,x^5 + f_6\,x^6
 + (3\,a\,x^2)^2y + g_5\,x^5y
 + h_3\,x^3y^2 + h_4\,x^4y^2
 + n_3\,x^3y^3
 + a_{24}\,x^2y^4 + a_{05}\,y^5 + a_{15}\,xy^5 + a_{06}\,y^6
$,
and $\dsp X=\pderiv{W}{x}$\,, $\dsp Y=\pderiv{W}{y}$\,, 
where the coefficients are non-negative constants,
 with $a>0$,
 such that $X^{2} (x,x^{2})-Y(x,x^{2})$ is a polynomial of $x$ with
non-negative coefficients.

Examples of the $2$ dimensional map 
$\Phi:\ (x,y)\mapsto (X(x,y),Y(x,y))$
satisfying the conditions
are the renormalization group (RG) map (modulo change of variables)
 for the restricted
 self-avoiding paths on the $3$ and $4$ dimensional pre-gaskets.

We prove that there exists a unique fixed point $(x_f,y_f)$ of $\Phi$
 in the invariant set
$\{(x,y)\in\preals^2\mid x^2\ge y\}\setminus\{0\}$.



\newpage

\section{Introduction and main results.}
\seca{intro}

In this paper, we study existence and uniqueness of fixed point for
a $2$ dimensional discrete time dynamical system in the first quadrant 
$\preals^2$, generated by the gradient
\eqnb
\eqna{PhiXY}
\Phi=(X,Y)=\grad W=(\pderiv{W}{x}\,, \pderiv{W}{y}):
\ \preals^2\to\preals^2
\eqne
of a polynomial $W:\ \preals^2\to\preals$ with non-negative coefficients,
such that the set
\eqnb
\eqna{Xi}
\Xi=\{(x,y)\in\preals^2 \mid y\le x^2\}
\eqne
is an invariant set of $\Phi=(X,Y)$.

Let us state our main results. The first result 
deals with the existence of fixed point in the interior of $\Xi$.
\thmb
\thma{existence}
Assume that $W:\ \reals^2\to\reals$ satisfies the following:
\itmb
\item
 $W$ is a polynomial in $2$ variables $x$ and $y$,
each term of which has positive coefficient and of total degree $3$ or more.
Moreover, the term $x^3$ exists (i.e., the coefficient of $x^3$ is
non-zero).

\item
$\Xi$ of \eqnu{Xi} is an invariant set of $\Phi=\grad W$:
if $(x,y)\in \Xi$, then $\Phi(x,y)\in\Xi$.
Moreover, $Y(x,x^2)<X(x,x^2)^2$, $x>0$, for $\Phi=(X,Y)$.

\item 
There exsits a term of the form $x^ny$ in $W$,
i.e., the coefficient of $x^ny$ is non-zero for some $n\ge 2$.

\item 
$R(x,z)=X(x,x^{2} z)^{2} -Y(x,x^{2} z)$ is a polynomial in
$z$, $1-z$, and $x$, with non-negative coefficients.
Namely, there exists a polynomial $\tilde{R}(x,z, s)$ in $3$ variables
with non-negative coefficients such that $R(x,z)=\tilde{R}(x,z,1-z)$.
Moreover, 
\eqnb
\eqna{Rsmall}
 \frac{R(x,z)}{Y(x,x^{2} z)} = O(x),\ \ x\to 0,
\eqne
where $O(x)$ is uniform in $z\in [0,1]$.

\itme
Then there exists a fixed point $(x_f,y_f)$ of $\Phi$ in the interior
$\interior{\Xi}=\{(x,y)\in\reals^2\mid x> 0,\ y>0,\ x^2> y\}$ of $\Xi$.
\DDD
\thme
We note that \thmu{existence} is not a direct consequence of
standard topological fixed point theorems on $\Xi$,
which allows for a fixed point on the boundary of $\Xi$,
$\partial \Xi=\{(x,0)\mid x\ge 0\}\cup\{(x,x^2)\mid x\ge 0\}$,
which is trivial, because $(0,0)$ is a fixed point of $\Phi$
under the conditions in the Theorem.
We are looking for a fixed point in $\interior{\Xi}$,
 the interior of $\Xi$, not on the boundary. 

We also note that restricting our attention to the subset $\Xi\subset\preals^2$
is essential, because outside $\Xi$, fixed points may dissappear and appear
with small changes in the coefficients of $W$.
For example, let $\dsp W_{\eps}(x,y) = \frac13 x^3 + x^4y + \eps y^6$.
(This choice satisfies the conditions in \thmu{existence} and \thmu{uniqueness}
below for $0\le\eps\le8/3$.)
Then for positive $\eps$, there are $4$ fixed points 
of $\Phi_{\eps}=\grad W_{\eps}$
in $\preals^2$; $(0,0)$,
 $(x_1,y_1)=(0.662\cdots +O(\eps),0.192\cdots +O(\eps))$,
$(0,(6\eps)^{-1/4})$, and one of order $(O(\eps^{1/8}), O(\eps^{-1/4}))$,
while for $\eps=0$ the last $2$ are absent and we have only $2$ fixed points.

An intuition for the specific conditions on $W$ in \thmu{existence}
arises in an attempt to extend a corresponding simple fact for function
with $1$ variable. Let $f(x)$ be a polynomial with non-negative coefficients
with lowest order term of $x^3$. Then there is a unique fixed point $x_f$
 of $f'$ on the positive $x$ axis ($f'(x_f)=x_f>0$).
Note that this is not the direct consequence of a standard topological
fixed point theorem on an obvious invariant set $\preals=\{x\ge 0\}$
 of the map $f'$, because $x=0$ is a fixed point. Rather, the unique
existence of the fixed point $x_f>0$ is due to the positivity of
 the coefficients in the map
and that a term $x^n$ with $n>1$ is smaller than $x$ for small $x$ and
 larger for large $x$.

 A simple way of extending this fact to $2$ variables
would be to assume that the second variable $y$ is of order $x^2$, at least
for small $x$, and that this relation is preserved under the map $\Phi=(X,Y)$
in consideration. This motivates the non-negativity of the coefficients
and conditions on $R=X^2-Y$ in \thmu{existence}.
We have added a couple of conditions to exclude fixed points on the boundary
on $\partial\Xi\setminus\{0\}$ to avoid complications.

\smallskip\par
We turn to our second result, which is on the
uniqueness of the fixed point $(x_f,y_f)$ of $\Phi$ in
 $\interior{\Xi}$.
This is a more difficult problem than the existence result, and we 
 have results only with $12$ adjustable coefficients for $W$,
in contrast to \thmu{existence} which allows for indefinitely many
terms.
\thmb
\thma{uniqueness}
Let $W:\ \preals^2\to\preals$ be a polynomial defined by
\eqnb
\eqna{W6} \arrb{l}\dsp
W(x,y) = a\,x^3 + b\,x^4 + f_5\,x^5 + f_6\,x^6
 + (3\,a\,x^2)^2y + g_5\,x^5y
 + h_3\,x^3y^2 + h_4\,x^4y^2
\\ \dsp \phantom{W(x,y) =}
 + n_3\,x^3y^3
 + a_{24}\,x^2y^4 + a_{05}\,y^5 + a_{15}\,xy^5 + a_{06}\,y^6,
\arre
\eqne
where all the constants
 $a,b,f_5,f_6,g_5,h_3,h_4,n_3,a_{24},a_{05},a_{15},a_{06}$,
are non-negative, and $a>0$,
and
$R(x,z)=X(x,x^{2} z)^{2} -Y(x,x^{2} z)$
is a polynomial in
$z$, $1-z$, and $x$, with non-negative coefficients,
in the same sense as in the corresponding condition in \thmu{existence}.
Then there exists a unique fixed point $(x_f,y_f)$ of
$\Phi=\grad W$ in $\interior{\Xi}$.
\DDD\thme
The condition on $R$ in \thmu{uniqueness} can be made explicit.
\prpb
\prpa{R510}
The conditions on $W$ in \thmu{uniqueness} is equivalent to the following:
$W$ is as in \eqnu{W6}, with the coefficients being non-negative,
$a>0$, and, $R_n\ge0$, $5\le n\le 10$, where $R_n$s are
\[\arrb{l}\dsp
R_5=  24\, a \,b - g_5 - 2\, h_3\,,
\\ \dsp
R_6=  16\, b^2 + 30\, a\, f_5 - 2\, h_4\,,
\\ \dsp
R_7=  216\, a^3 + 40\, b\, f_5 + 36\, a\, f_6 - 3\, n_3\,,
\\ \dsp
R_8=  288\, a^2\, b + 25\, f_5^2 + 48\, b\, f_6 + 30\, a\, g_5 + 
      18\, a\, h_3
 - 5\, a_{05} - 4\, a_{24}\,,
\\ \dsp
R_9=  360\, a^2\, f_5 + 60\, f_5\, f_6 + 40\, b\, g_5 + 
      24\, b\, h_3 + 24\, a\, h_4 - 5\, a_{15}\,,
\\ \dsp
R_{10}= 648\, a^4 + 216\, a^2\, f_6 + 18\, f_6^2 + 25\, f_5\, g_5 + 
   15\, f_5\, h_3 + 16\, b\, h_4
 + 9\, a\, n_3 - 3\, a_{06}\,.
\ \ \ \mbox{\DDD}\arre \]
\prpe
That this is necessary is easily seen, if one explicitly writes
the coefficients of $x^n$ in $R(x,1)$ for $5\le n\le 10$.
That the conditions in \prpu{R510} are sufficient is proved by
looking into the coefficients of $x^n$ in $R(x,z)$ (each of which is
a polynomial in $z$).
It turns out that with $W$ of the form \eqnu{W6}, terms with $x^n$ appear
for $5\le n\le 20$, among which no explicit negative signs appear
 for $n\ge 11$, hence the condition hold automatically,
 and for the remaining $5\le n\le 10$,
the power of $z$ in the terms with negative signs are larger than any of
the terms with positive signs, hence with the non-negativity conditions
 at $z=1$, assumed in \prpu{R510},
it is straightforward to find a polynomial in $z$ and $1-z$ with
non-negative coefficients. \prpu{R510} is thus proved.

Among the examples of $W$ satisfying the conditions
in \thmu{uniqueness}, or equivalently, in \prpu{R510},
are those related to the renormalization group (RG) map for the restricted
 self-avoiding paths on the $3$ and $4$ dimensional pre-gaskets
\cite{HHK93,dSG01,kyoritu}:
\eqnb
\eqna{SAPdSG}
\arrb{l}\dsp
W_3(x,y)=
\frac13\,x^3+\frac12\,x^4+\frac{2}{5}\,x^5+x^4y+2\,x^3y^2+\frac{22}{5}\,y^5,
\\ \dsp
W_4(x,y)=
\frac{\sqrt{3}}{9}\, x^3 + \frac14\,x^4 + \frac{2\sqrt{3}}{15}\, x^5
 + \frac{1}{9}\, x^6 + \frac{1}{3}\, x^4 y
\\ \dsp{}
 + \frac{2\sqrt{3}}{9}\, x^5 y
 + \frac{2\sqrt{3}}{9}\, x^3 y^2
 +\frac{13}{18}\, x^4 y^2 + \frac{32\sqrt{3}}{81}\, x^3 y^3
 +\frac{22}{27}\, x^2 y^4
\\ \dsp{}
 + \frac{22}{135}\, y^5
 +\frac{44\sqrt{3}}{81}\, x y^5
 + \frac{31}{81}\, y^6.
\arre
\eqne
It is straightforward to see that $W_3$ and $W_4$ satisfy all the conditions
in \prpu{R510}.
The fixed point equation $(x,y)=\Phi(x,y)$ for $\Phi=\grad W_3$ is
\[ \arrb{l} \dsp
x=x^2+2x^3+2x^4+4x^3y +6x^2y^2,
\\ \dsp
y=x^4+4x^3y+22y^4,
\arre \]
which coincides with that for $\vec{\Phi}$
 in \cite[(2.3) and (2.4)]{HHK93},
and
the fixed point equation $(x,y)=\Phi(x,y)$ for $\Phi=\grad W_4$ is,
with the change of variables $x=\sqrt{3}\,x'$ and $y=3y'$,
\[ \arrb{l} \dsp
x'=x'^2+3{x'}^3+6{x'}^4+ 6{x'}^5+12{x'}^3y'+30{x'}^4y'+18{x'}^2{y'}^2
\\ \dsp \phantom{x'=}
+78{x'}^3{y'}^2+96{x'}^2{y'}^3+132{x'}{y'}^4 + 132{y'}^5,
\\ \dsp
y'={x'}^4+2{x'}^5+4{x'}^3y'+13{x'}^4y'+32{x'}^3{y'}^2+88{x'}^2{y'}^3
+22{y'}^4+220{x'}{y'}^4+186{y'}^5,
\arre \]
which coincides with the fixed point equation for $\vec{\Phi}$
 in \cite[(33)]{dSG01} with $x\mapsto x'$ and $y\mapsto y'$.
A motivation of the conditions in \thmu{uniqueness} was an attempt
to generalize the known examples \eqnu{SAPdSG}.

The class of $W$ allowed by the conditions in \thmu{uniqueness} is 
a subset of that in \thmu{existence}.
This may be easily seen from the following equivalent conditions 
to those in \thmu{uniqueness}.
\prpb
\prpa{fromgasket}
The conditions on $W$ in \thmu{uniqueness} is equivalent to the following:
\itmb
\item
The conditions in \thmu{existence} hold.
\item
Each term has total degree no more than $6$.
\item
Terms containing positive powers of $y$ has total degree $5$ or $6$.
\item
$xy^4$ and $x^2y^3$ are absent.
\DDD
\itme
\prpe
That these conditions imply those in \thmu{uniqueness} is easily seen,
if one notices that the extra conditions in \prpu{fromgasket} implies
\eqnu{W6} modulo the coefficient of $x^4y$, which is fixed by the condition
$R(x,z)/Y(x,x^2z)=O(x)$. The converse is proved in a similar way.

In \cite{HHK93} and \cite{dSG01}, the results in \thmu{existence} and
\thmu{uniqueness} are proved for 
$W=W_3$ and $W=W_4$ in \eqnu{SAPdSG}, respectively, but the proofs there
explicitly uses the explicit values of coefficients in $W_3$ and $W_4$.
These values of the coefficients are essentially the numbers of certain figures
(self-avoiding paths) on the $3$ and $4$ dimensional gaskets, respectively,
and the non-Markovian nature of self-avoiding paths makes it hard
to count these numbers, not to mention to find general formula for all $d$
dimensional gaskets.
It is therefore important for the RG approach that the results in the above
theorems could be derived from `basic properties' for $W$ which can be
derived by simple arguments.
It is not very difficult to derive the conditions in \prpu{fromgasket}
 (including those in \thmu{existence}) from basic graphical considerations
in the case of restricted self-avoiding paths on $3$ and $4$ dimensional
gaskets,
so \thmu{uniqueness} provides a rather satisfactory alternate proofs
to the corresponding original proofs in \cite{HHK93,dSG01},
in that one no more needs to count the number of self-avoiding paths exactly,
for a proof of existence and uniqueness of fixed points in $\Xi^o$.

We also note that the examples \eqnu{SAPdSG} do not seem to fit to
any existing general theorems on fixed point uniqueness,
much less the class in \thmu{uniqueness}. This may reflect the fact that
self-avoiding paths are non-Markovian and mathematically hard to analyze.
The present study may then provide a new direction in the study of
fixed point theorems.

\smallskip\par
A plan of this paper is as follows.
In \secu{existence} we prove \thmu{existence} and
in \secu{uniqueness} we prove \thmu{uniqueness}.

We note that our proof for \thmu{uniqueness} in \secu{uniqueness}
in fact proves a stronger property than is stated in the Theorem,
and does not hold for all $W$ in the class satisfying the conditions
in \thmu{existence}. However, it seems that even for examples where the
proof in \secu{uniqueness} breaks down, the statement in \thmu{uniqueness}
still seems to hold. We therefore close this Introduction with the following
conjecture.
\cnjb
Uniqueness of the fixed point $(x_f,y_f)$ of $\Phi=\grad W$
 in $\interior{\Xi}$ hold under the
condition in \thmu{existence}.
\DDD\cnje

 \bigskip\par \noindent{\bf Acknowledgements.}

The author would like to thank Prof.~Hidekazu Ito for
suggestions and his interest in the present work.

The research is supported in part by a Grant-in-Aid for 
Scientific Research (B) 17340022 from the Ministry of Education,
 Culture, Sports, Science and Technology.

\section{Proof of existence of fixed point.}
\seca{existence}

Here we prove \thmu{existence}.

Let $W=W(x,y)$ be a polynomial with non-negative coefficients
satisfying the conditons in \thmu{existence}, and
define functions in $2$ variables $F$ and $G$ by
\eqnb
\eqna{G}
 G(x,z) = \frac{1}{x} X(x,x^2z),
\eqne
and
\eqnb
\eqna{F}
F(x,z) = z \frac{X^{2}(x,x^2z)}{Y(x,x^2z)}\,,
\eqne
where $\grad W=(X,Y)$.

Note that
with the change of variables $(x,y)\mapsto (x,z)$ defined by $y=x^2z$,
the set
 $\Xi\setminus\{0\}=\{(x,y)\in\preals^2 \mid y\le x^2\}\setminus\{0\}$
is mapped to a strip in the first quadrant $\preals^2$
\eqnb
\eqna{tildeXi}
\tilde{\Xi}=\{(x,z)\in \reals^2\mid x>0,\ 0\le z\le 1\}.
\eqne
Since $W(x,y)$ has a term of a form $x^ny$ by
a condition in \thmu{existence}, $Y(x,x^2z)>0$, $(x,z)\in \tilde{\Xi}$,
hence $F(x,z)$ is well-defined, positive and analytic on $\tilde{\Xi}$.

Note also that
 $(x,x^2z)\in\Xi\setminus\{0\}$ is a fixed point of $\Phi=(X,Y)$
if and only if
 $(x,z)\in\tilde{\Xi}$ and $F(x,z)=G(x,z)=1$.
\lemb
\lema{existence}
$G(x,z)$ is a polynomial in $x$ and $z$ with non-negative coefficients,
satisfying
\eqnb
\eqna{Gsmall}
G(x,z)=3ax\, (1+O(x)),
\eqne
 where
$a$ is the coefficient of the term $x^3$ in $W$.
Furthermore, the contour set for $G=1$ in the strip $\tilde{\Xi}$ is
a smooth curve connecting the floor $z=0$ and the ceiling $z=1$;
Namely, there exists a positive continously differentiable
function $x^*(z)>0$ for $0\le z\le 1$ such that 
\eqnb
\eqna{Gcontour}
\{(x,z)\in\tilde{\Xi}\mid G(x,z)=1\}=
\{(x^*(z),z)\mid 0\le z\le 1\}.
\eqne

$F$ satisfies $F(x,z)>0$, $(x,z)\in \interior{\tilde{\Xi}}$,
$F(x,0)=0$ and
$F(x,1)>1$ for $x>0$.
\DDD\leme
\prfb
All the statement about $G$ is obvious from the conditions
 in \thmu{existence}, except perhaps the last one.
To see that the stated $x^*$ exists,
first note that by the conditions in \thmu{existence},
 $W$ is a polynomial with non-negative coefficients
with lowest order being $x^3$, hence $G$ is a polynomial 
with non-negative coefficients satisfying
 $\dsp \pderiv{G}{x}(x,z)\ge 3a >0$,
 $\dsp\lim_{x\downarrow 0} G(x,z)=0$, and 
 $\limf{x} G(x,z)=\infty$ for $0\le z\le 1$.
This with an implicit function theorem implies that
there uniquely exists a continuously differentiable function
$x^*:\ [0,1]\to \reals_{>0}$ such that
$G(x^*(z),z)=1$, $0\le z\le 1$,
and monotonicity of $G$ in $x$ implies that all the point 
satisfying $G(x,z)=1$ is on the curve $\{(x^*(z),z)\}$.

Statements on $F$ are also easy, if one notes
\eqnb
\eqna{existenceprf}
 F(x,z)=z\,\left(1+\frac{R(x,z)}{Y(x,x^2z)}\right).
\eqne
\QED
\prfe

A proof of \thmu{existence} is now obvious,
because \lemu{existence} implies
$F(x^*(0),0)=0$ and $F(x^*(1),1)>1$, 
for a smooth curve
 $\{(x^*(z),z)\mid 0\le z\le 1\} \subset\tilde{\Xi}$ hence
there is a $z^*\in (0,1)$ such that $F(x^*(z^*),z^*)=G(x^*(z^*),z^*)=1$,
which, as noted at the beginning of this section, implies the existence
of a fixed point of $\grad W=(X,Y)$ in $\interior{\Xi}$.
\QED

\section{Proof of uniqueness of fixed point.}
\seca{uniqueness}

Here we prove \thmu{uniqueness}.

Let $J_{GF}(x,z)$ be the Jacobian matrix of the map
 $(x,z)\mapsto (G(x,z),F(x,z))$;
\eqnb
\eqna{JGF}
 J_{GF}=
\pderiv{G}{x} \pderiv{F}{z} - \pderiv{F}{x} \pderiv{G}{z}\,.
\eqne
A core of our proof of \thmu{uniqueness}
is to prove $J_{GF}\ne 0$ on the contour curve $G=1$ in $\tilde{\Xi}$.
This implies that the map is locally one-to-one,
which further implies, with additional properties such as
\eqnu{Gcontour} and \lemu{preliminary} below,
 global one-to-one properties, implying uniqueness of the fixed point.

The proof of \thmu{uniqueness} in this section starts with
\lemu{preliminary}, then follows \lemu{good} where
we prove that, with these properties,
positivity of Jacobian $J_{GF}$ is sufficient for a proof of \thmu{uniqueness}.
Up to this pont, the arguments are `soft' and
 all the results hold for the class in \thmu{existence}.
The hardest part comes last, a proof
that $J_{GF}>0$ for the class assumed in \thmu{uniqueness}.
That this is hard may be seen if one notices that
outside $\tilde{\Xi}$ there may be more than one fixed points
(as are the cases for the examples below \thmu{existence} and \eqnu{SAPdSG}
\cite{HHK93,dSG01}), hence $J_{GF}<0$ actually occurs for some
$(x,z)\in \preals^2$.
We must therefore find a nice quantity which is explicitly positive only in
a subset of $\tilde{\Xi}$ and then prove (as we will in \lemu{Rem0})
 that the quantity is a lower bound
of $J_{GF}$ using inequalities in \prpu{R510}.

\lemb
\lema{preliminary}
Under the conditions in \thmu{existence},
if $x>0$ is sufficiently small, then 
\[ J_{GF}(x,z)>0,\ \ 0\le z\le 1, \]
 and
furthermore, the set of $(x,z)\in \tilde{\Xi}$ satisfying $F(x,z)=1$
is a single curve for small $x$,
having $(0,1)$ as an endpoint.
(More precisely, there exists $\delta>0$ such that
$\{(x,z)\in (0,\delta]\times [0,1]\mid F(x,z)=1\}$
is a curve whose endpoints are $(0,1)$ and a point on $x=\delta$.)
\DDD\leme
\prfb
\eqnu{existenceprf} and \eqnu{Rsmall} imply
\eqnb
\eqna{Fsmall}
F(x,z)=z\, (1+O(x)),
\eqne
uniformly in $0\le z\le 1$,
which, with \eqnu{Gsmall} implies $J_{GF}=a + O(x)>0$
for small $x$, say $0<x<\delta$.
This in particular implies $\grad F\ne 0$, hence
$\{(x,z)\in (0,\delta]\times[0,1]\mid F(x,z)=1\}$
is a finite union of non-intersecting smooth curves,
each segment of which is either closed or with endpoints
at $x=0$ or $x=\delta$.

\lemu{existence} implies $F(x,1)>1$ and $F(x,0)=0$,
so that $F=1$ cannot intersect $z=0$ nor $z=1$.
Also \eqnu{Fsmall} implies $F(+0,z) = z$,
which, with $F(x,1)>1$ implies that
the contour curve for $F=1$ exists and
intersect $x=0$ at $z=1$.
By definition,
\[ \arrb{l}\dsp
F(x,1-u)-1 =(1-u)\,\frac{X^2}{Y}(x,x^2\,(1-u))-1
=\frac{R(x,1-u)-uX^2(x,x^2\,(1-u))}{Y(x,x^2\,(1-u))}\,, 
\arre \]
and
\[ \arrb{l}\dsp
R(x,1-u)-uX^2(x,x^2\,(1-u))
=R(x,1) - 9a^2x^4\,(1+O(x)) u +x^7\,O_x(u^2)
\\ \dsp {}
=O(x^5)- 9a^2x^4\,(1+O(x)) u.
\arre \]
Hence, For small $x$ and $u$,
the contour $F(x,1-u)=1$ is uniquely given by
$u= O(x)$ in $(x,1-u)\in (0,\delta]\times[0,1]$.
\QED\prfe

\lemb
\lema{good}
In addition to the conditions in \thmu{existence}, assume that
$J_{GF}(x,z)\ne 0$ on
\eqnb
\eqna{Xip}  
 \tilde{\Xi}'
=\{ (x,z)\in (0,\infty)\times (0,1)\mid  G(x,z)\le 1 ,\ F(x,z)\le 1\}
\ \ \  \subset \interior{\tilde{\Xi}}.
\eqne
Then the fixed point of $\Phi=\grad W =(X,Y)$ is unique in $\interior{\Xi}$.
\DDD\leme
\prfb
As noted at the beginning of \secu{existence},
$(x,y)\in\interior{\Xi}$ is a fixed point of $\Phi$ if and only if
$(x,y/x^2)\in\interior{\tilde{\Xi}}$ and $F(x,y/x^2)=G(x,y/x^2)=1$.
\thmu{existence} implies that there is a fixed point
 $(x_f,y_f)\in\interior{\Xi}$ of $\Phi$.
Put $z_f=y_f/x_f^2$. 

Then $F(x_f,z_f)=G(x_f,z_f)=1$,
hence $(x_f,z_f)\in \tilde{\Xi}'$.

As in the proof of \lemu{preliminary},
$J_{GF}\ne 0$ implies $\grad F\ne 0$, which further implies
that
\[ A=\{(x,z)\in \tilde{\Xi}' \mid F(x,z)=1\}\ \ \subset \tilde{\Xi}' \]
is a finite union of non-intersecting smooth curves,
each segment of which is either closed, or with one endpoint
 $(0,1)$ and the other on $\{(x,z)\in \tilde{\Xi}'\mid G(x,z)=1\}$.
 $(x_f,z_f)$ is on one of such curves.

Suppose that $(x_f,z_f)$ is on a smooth curve
 $C:\ [0,1]\to A$
with $\dsp \diff{C}{s}(s)\ne 0$, $0<s<1$,
whose endpoints are both on $G=1$; $G(C(0))=G(C(1))=1$ and
$C(s)\in \tilde{\Xi}'$, $0\le s\le 1$.
A mean-value Theorem then implies that there exists $s_0\in(0,1)$
such that
\[ 0=\left. \diff{G(C(s))}{ds}\right|= (\grad G)(C(s_0)) \cdot
\diff{C}{s}(s_0). \]
Multiplying by $\dsp (\pderiv{F}{z}\,,\ -\pderiv{F}{x})(C(s_0))$
from left, we have
$\dsp  0=J_{GF}(C(s_0)) \diff{C}{s}(s_0), $
which contradicts the assumption $J_{GF}\ne 0$ on $\tilde{\Xi}'$.
Therefore, a contour curve in A on which $(x_f,z_f)$ exists,
cannot have both endpoints on $G=1$.
Similarly, such a curve cannot be a closed curve in $\tilde{\Xi}'$.
Therefore the curve must have one endpoint $(0,1)$ and 
the other on $G=1$, the latter endpoint being the point $(x_f,z_f)$.

By \lemu{preliminary}, a curve of the contour set $F=1$
that has endpoint $(0,1)$ is unique.
Therefore there is only one curve $C\subset A \subset \tilde{\Xi}'$
 on which there is a point satisfying $F(x_f,z_f)=G(x_f,z_f)=1$,
hence, as noted at the beginning of the proof,
the fixed point $(x_f,x_f^2z_f)$ is unique in $\interior{\Xi}$.
\QED
\prfe

A proof of \thmu{uniqueness} is now reduced to proving
$J_{GF}(x,z)\ne 0$ on \eqnu{Xip} under the conditions in \thmu{uniqueness}.
This follows as the direct consequence of the following \lemu{Rem0}.
In fact, the Lemma states positivity of $J_{GF}(x,z)$ on 
\eqnb
\eqna{Xipp}  
 \tilde{\Xi}''
=\{ (x,z)\in (0,\infty)\times (0,1)\mid  F(x,z)\le 1\}
\ \ \  \subset \interior{\tilde{\Xi}},
\eqne
which is larger than \eqnu{Xip}.
\lemb
\lema{Rem0}
Assume that $W$ satisfies the conditions in \thmu{uniqueness}.
Let $e$ be a function defined by
\eqnb
\eqna{Rem0}  
 e(x,z)=(1-z)x^2 \frac{Y^2}{X^2}(x,x^2z)\,\left(
J_{GF}-\frac{F(1-F)}{z(1-z)}\, \pderiv{G}{x}\right)(x,z).
 \eqne
Then $e(x,z)$ is a polynomial in $x$, $z$, $1-z$ with non-negative
 coefficients.
Namely, there exists a polynomial $f(x,z, s)$ in $3$ variables
with non-negative coefficients such that
$e(x,z)=f(x,z,1-z)$.
Furthermore, $f(x,z, s)$ has a term $a^4\, s^2zx^9$,
hence in particular, $e(x,z)>0$, and consequently, $J_{GF}(x,z)>0$,
$(x,z)\in \tilde{\Xi}''$.
\DDD
\leme
\prfb
The last claim is by explicit calculation of order $x^{9}$ terms.
For this and for the calculations below, we use Mathematica software
to assist the simple algebraic manupulation such as expanding
and factoring. ($e(x,z)$ has more than $300$ terms with positive
coefficients and more than $80$ terms with negative ones!)

The problem is to use $R_n$s in \prpu{R510} to eliminate
apparent negative signs in $e(x,z)$.
It turns out that we have an expression
\eqnb
\eqna{proofofJ6}
e(x,z)=e_c(x,z,1-z)   
+e_r(x,z,1-z)         
\eqne
where
\begin{flushleft}
$
e_c(x,z, s ) =
\ 3\, a\, R_5\, z\, x^7
\ +\ 3\, a\, R_6\, z\, x^8
\ +\ 8\, b\, R_5\, z\, x^8
\ +\ 3\, a\, R_7\, z^2\, (1 +  s )\, x^9
\ +\ 8\, b\, R_6\, z\, x^9
\ +\ 15\, f_5\, R_5\, z\, x^9
\ +\ 3\, a\, R_8\, z^3\, (1 + 2\,  s )\, x^{10}
\ +\ 8\, b\, R_7\, z^2\, (1 +  s )\, x^{10}
\ +\ 15\, f_5\, R_6\, z\, x^{10}
\ +\ R_5\, (a^2\, (144\, z^3 + 36\, z^3\,  s ) + f_6\, 24\, z)\, x^{10}
\ +\ 8\, b\, R_8\, z^3\, (1 + 2\,  s )\, x^{11}
\ +\ 15\, f_5\, R_7\, z^2\, (1 +  s )\, x^{11}
\ +\ R_6\, (6\, a^2\, z^3\, (18 + 6\, z + 6\, z\,  s ) + 
 f_6\, 24\, z)\, x^{11}
\ +\ R_5\, (g_5\, (3\, z^2 + 45\, z^2\,  s  + 6\, z^4 + 16\, z^5) + 
 h_3\, (4\, z^2 + 12\, z^2\,  s ^2 + 11\, z^5) + 
 a\, b\, 24\ \,  s ^2\, z^2\, (8 + z))\, x^{11}
\ +\ R_9\, (3\, z^4 + 9\, z^4\,  s )\, a\, x^{11}
\ +\ R_{10}\, a\, 6\, \ (1 + 4\,  s )\, z^5\, x^{12}
\ +\ R_9\, b\, 8\, (1 + 3\,  s )\, z^4\, x^{12}
\ +\ R_7\, f_6\, 24\, (1 +  s )\, z^2\, x^{12}
\ +\ R_8\, f_5\, (1 + 2\,  s )\, z^3\, 15\, x^{12}
\ +\ R_7\, a^2\, 18\, z^3\, (3 + 4\, z + z^2 + 4\, z^2\,  s )\, x^{12}
\ +\ R_6\, h_3\, (6\, z^4 + 9\, z^5)\, x^{12}
\ +\ R_6\, g_5\, 5\, z^2\, (9\,  s  + z^2 + 5\, z^2\,  s  +
 4\, z^4)\, x^{12}
\ +\ R_5\, h_4\, 4\, z^3\, (1 + 8\,  s  + z^2 + 4\, z^3)\, x^{12}
\ +\ R_9\, f_5\, 15\, (1 + 3\,  s )\, z^4\, x^{13}
\ +\ R_{10}\, b\, 16\, (1 + 4\,  s )\, z^5\, x^{13}
\ +\ R_8\, f_6\, 24\, z^3\, (1 + 2\,  s )\, x^{13}
\ +\ R_8\, a^2\, 9\, z^4\, (16 + 4\, (1 + z)\,  s )\, x^{13}
\ +\ R_7\, g_5\, (25\, z^3\,  s  + 25\, z^3\, (1 + z)\,  s  +
 25\, z^6)\, x^{13}
\ +\ R_5\, n_3\, 3\, z^3\, (10\, z\,  s  + 7\, z^3)\, x^{13}
\ +\ R_7\, h_3\, (5\, z^5 + 10\, z^6)\, x^{13}
\ +\ R_5\, a\, f_6\, 108\ z^3\,  s ^3\, x^{13}
\ +\ R_6\, h_4\, (16\, z^4 + 8\, z^4\, (1 + z)\,  s  + 8\, z^7)\, x^{13}
\ +\ R_{10}\, f_5\, 30\, z^5\, (1 + 4\,  s )\, x^{14}
\ +\ R_9\, f_6\, 24\, (1 + 3\,  s )\, z^4\, x^{14}
\ +\ R_9\, a^2\, 18/5\, (39 + 46\,  s )\, z^5\, x^{14}
\ +\ R_6\, n_3\, (21\, z^5 + 3\, z^6)\, x^{14}
\ +\ R_7\, h_4\, 22\, z^5\, x^{14}
\ +\ R_8\, g_5\, (25\, z^4 + 8\, z^4\,  s  +
 6\, z^4\, (1 + z)\,  s )\, x^{14}
\ +\ R_8\, h_3\, 3\, (5 + 3\ \,  s )\, z^6\, x^{14}
\ +\ R_5\, a_{24}\, (10\, z^5 + 6\, z^6)\, x^{14}
\ +\ R_{10}\, f_6\, 48\, z^5\, (1 + 4\,  s )\, x^{15}
\ +\ R_{10}\, a^2\, 288\, (1 + 2\,  s )\, z^6\, x^{15}
\ +\ R_9\, h_3\, 18\, z^5\, x^{15}
\ +\ R_5\, a_{15}\, 12\, z^6\, x^{15}
\ +\ R_9\, g_5\, 2/5\, (61 + 94\,  s )\, z^5\, x^{15}
\ +\ R_6\, a_{24}\, 24\, z^5\, x^{15}
\ +\ R_8\, h_4\, z^5\, (4 + 20\, z + 20\, z\,  s )\, x^{15}
\ +\ R_7\, n_3\, 21\, z^6\, x^{15}
\ +\ R_8\, n_3\, z^7\, (21 + 9\,  s )\, x^{16}
\ +\ R_{10}\, g_5\, (50\, z^6 + 120\, z^6\,  s )\, x^{16}
\ +\ R_{10}\, h_3\, (30\, z^7 + 24\, z^7\,  s )\, x^{16}
\ +\ R_9\, h_4\, (23\, z^6 + 8\, z^6\,  s )\, x^{16}
\ +\ R_7\, a_{24}\, 26\, z^6\, x^{16}
\ +\ R_6\, a_{15}\, 13\, z^6\, x^{16}
\ +\ R_{10}\, h_4\, (48\, z^7 + 64\, z^7\,  s )\, x^{17}
\ +\ R_9\, n_3\, (17\, z^7 + z^8\ + 7\, z^8\,  s )\, x^{17}
\ +\ R_8\, a_{24}\, 16\, z^8\, x^{17}
\ +\ R_7\, a_{15}\, (2\, z^8 + 12\, z^9)\, x^{17}
\ +\ R_{10}\, n_3\, (42\, z^8 + 24\, z^8\,  s )\, x^{18}
\ +\ R_9\, a_{24}\, (3\, z^8 + 14\, z^9)\, x^{18}
\ +\ R_8\, a_{15}\, (10\, z^8 + 5\, z^9)\, x^{18}
\ +\ R_{10}\, a_{24}\, 32\, z^9\, x^{19}
\ +\ R_9\, a_{15}\, (z^9 + 8\, z^{10})\, x^{19}
\ +\ R_{10}\, a_{15}\, (10\, z^{10} + 8\, z^{11})\, x^{20}
 \,,$
\end{flushleft}
with which the remainder $e(x,z)-e_c(x,z,1-z)$ has an expression
$e_r(x,z,1-z)$ for a polynomial $e_r(x,z, s)$ with
non-negative coefficients, where non-negativity of $e_r$
holds by non-negativity of $a,b,\cdots$, without the conditions $R_n\ge 0$.
For completeness, we give an explicit form of $e_r$ in the appendix.

By \eqnu{proofofJ6}, $e(x,z)=f(x,z,1-z)$ with
$\dsp f(x,z,s)=e_c(x,z,s)+e_r(x,z,s)$, which
completes a proof of \lemu{Rem0}, hence, as noted at the beginning
of this section, a proof of \thmu{uniqueness} is also complete.
\QED\prfe

We remark that \lemu{Rem0} is a result much stronger than
is required to prove \thmu{uniqueness}.
In fact, with \lemu{Rem0}, a similar argument as for the 
contour curves $F=1$ and $G=1$ hold for contours $G=c$ for
any $c>0$ and $F=c$ for any $c\le 1$, hence in particular,
we have the following.
\corb
\cora{Fle1}
$\tilde{\Xi}''$ defined by \eqnu{Xipp} is a connected set,
whose boundary is $\{x=0\}\cup\{z=0\}\cup\{F=1\}$, and
the map $(x,z)\mapsto (G,F)$ is globally one-to-one on $\tilde{\Xi}''$.
\DDD\core 

We also remark that the formula \eqnu{Rem0} and the rather lengthy $e_c$
was found to work by trial and error, and it is an open problem to find
their intuitive (either mathematical or physical) meaning.

\appendix
\section{An explicit form of $e_r$.}

For completeness, we will give an explicit form of $e_r$ defined in the 
proof of \lemu{Rem0}. (Note that it is not unique. For example,
there are more than $1$ ways of writing $3-2z$ as a polynomial of 
$z$ and $1-z$ with positive coefficients; $3-2z=1+2(1-z)=z+3(1-z)$.)

$\dsp e_r(x,z,s) = \sum_{n=9}^{30} C[n, z, s]\, x^n$, where,
\begin{flushleft}
\small
$C[9, z_, s_] = \,
      12\, a\, (54\, a^3 + 10\, b\, f_{5} + 9\, a\, f_{6})\, s^2\, z$,

$C[10, z_, s_] = \,
      s^2\, z\, (320\, b^2\, f_{5} + 
            144\, a\, b\, (18\, a^2\, z + f_{6}\, (3 + 2\, z)) + 
            3\, a\, (25\, f_{5}^2\, (1 + 2\, z) + 
     3\, (22\, a\, g_{5} + 16\, a\, g_{5}\, z + 4\, a\, h_{3}\, z +
                        5\, a_{05}\, z^2)))$,

$C[11, z_, s_] = \,
 s^2\, z\, (15\, g_{5}^2 + 38\, g_{5}^2\, z + 110\, g_{5}\, h_{3}\, z + 
 32\, h_{3}^2\, z + 
   16\, g_{5}^2\, z^2 + 43\, g_{5}\, h_{3}\, z^2 + 22\, h_{3}^2\, z^2 + 
            144\, a^2\, h_{4}\, z\, (1 + z) + 
            2160\, a^3\, f_{5}\, z\, (1 + 2\, z) + 
            384\, b^2\, (f_{6} + 2\, f_{6}\, z) + 
            180\, a\, f_{5}\, f_{6}\, (4 + 2\, z + 3\, z^2) + 
        40\, b\, (3\, a_{05}\, z^2 + 10\, f_{5}^2\, (2 + z)))$,

$C[12, z_, s_] = \,
      s^2\, z\, (225\, a_{05}\, f_{5}\, z^2 + 432\, b^2\, h_{3}\, z^2 + 
            360\, a\, f_{5}\, h_{3}\, z^3 + f_{5}^3\, (375 + 750\, z) + 
            b\, f_{5}\, f_{6}\, (2160 + 2400\, z + 1440\, z^2) + 
            b^2\, g_{5}\, (400\, z^2 + 320\, z^2\, s) + 
            h_{3}\, h_{4}\, (90\, z^2 + 32\, z^3) + 
            g_{5}\, h_{4}\, (104\, z + 66\, z^2 + 56\, z^3) + 
            a\, f_{6}^2\, (972 + 216\, z + 324\, z^2 + 432\, z^3) + 
            a^2\, b\, f_{5}\, (12960\, z + 10800\, z^2 + 2880\, z^3) + 
            a^3\, f_{6}\, (1296\, z + 5832\, z^2 + 7776\, z^3) + 
            a^5\, (23328\, z^2 + 31104\, z^3))$,

$C[13, z_, s_] = \, 
      z\, (72\, a\, b\, n_{3}\, z^5 + 
            a_{05}\, f_{6}\, 360\, s^2\, z^2 + 
            f_{5}^2\, f_{6}\, 300\, s^2\, (5 + 10\, z + 9\, z^2) + 
            a^3\, g_{5}\, 1080\, z^4\, (1 + 4\, s) + 
            a^3\, h_{3}\, 216\, z^3\, (3 + s\, (12 + 7\, z)) + 
            a\, f_{6}\, g_{5}\, 36\, z\, (\, 
             5\, z^4 + 25\, z^3\, s + 35\, s^2 + s^3\, (15 + 8\, z)) + 
            a\, f_{6}\, h_{3}\, 36\, z^3\, ((3 + 16\, z)\, s + 3) + 
            a^2\, a_{24}\, ( 36\, z^4 + 144\, z^4\, s) + 
            h_{3}\, n_{3}\, 72\, s^2\, z^3 + 
            g_{5}\, n_{3}\, 3\, s\, z^2\, ( 32\, s\, (1 + z) + 2 + 3\, z) + 
            b\, f_{5}\, h_{3}\, 80\, s\, z^2\, ( 5 + s\, (10 + 7\, z)) + 
            b\, f_{5}\, g_{5}\, 200\, s^2\, z\, (10 + 10\, z + 3\, z^2) + 
            a^2\, a_{05}\, 900\, s^2\, z^3 + 
            a^2\, f_{5}^2\, 900\, s^2\, z\, (10 + 20\, z + z^2) + 
            b\, f_{6}^2\, 288\, s^2\, (5 + 10\, z + 3\, z^2 + 4\, z^3) + 
            a^2\, b\, f_{6}\, 864\, s^2\, z\, (20 + 13\, z + 21\, z^2) + 
            a^4\, b\, 51840\, s^2\, z^2\, (1 + z) + 
            a\, f_{5}\, h_{4}\, 120\, z^2\, ( 1\,  + 2\, z^3\, s + 9\, s^2) + 
            b^2\, h_{4}\, 128\, s^2\, z^2\, (5 + 6\, z + z\, s) + 
            h_{4}^2\, 16\, s^2\, z^2\, (1 + z + z^2))$,

$C[14, z_, s_] = \, 
      z/5\, (a\, f_{6}\, h_{4}\, 360\, z^2\, ( 3 + 6\, s + 13\, s^2 )  + 
            a\, a_{24}\, b\, (720\, z^4 + 720\, z^4\, s) + 
            b^2\, n_{3}\, 240\, z^3\, ( 3 + s\, (8 + z)) + 
            a_{24}\, g_{5}\, 90\, z^4\, s + h_{4}\, n_{3}\, 30\, z^3\, s^2 + 
            a\, g_{5}^2\, 75\, z^2\, (5\, (1 + s)^2 + s^2\, (35 + 12\, z)) + 
            a_{05}\, g_{5}\, 975\, s^2\, z^3 + 
            b\, f_{6}\, g_{5}\, 240\, z\, (
                4\, z^3 + 6\, z^4 + 46\, s + (9 + 64\, z)\, s^2) + 
            a\, f_{5}\, n_{3}\, 4950\, s^2\, z^3 + 
   b\, f_{5}\, h_{4}\, 400\, z^2\, (
  3\, z^2 + 2\, z^3 + 22\, s + 22\, z\, s^2) + 
     f_{5}\, f_{6}^2\, 900\, s^2\, (11 + 22\, z + 33\, z^2 + 12\, z^3) + 
            f_{5}^2\, g_{5}\, 125\, s^2\, z\, (55 + 110\, z + 126\, z^2) + 
            a^3\, h_{4}\, 864\, z^3\, (8 + 24\, s + 23\, s^2) + 
            a^2\, b\, g_{5}\, 720\, z^2\, (
                21\, z^3 + 105\, s + (5 + 37\, z)\, s^2) + 
            a^2\, f_{5}\, f_{6}\, 1080\, z\, (
                z^4 + z^2\, 5\, s + 55\, s^2\, (2 + 4\, z + 3\, z^2)) + 
            a^4\, f_{5}\, 6480\, z^2\, (
                z^3 + z^2\, 5\, s + 55\, (1 + 2\, z)\, s^2) + 
            a_{24}\, h_{3}\, 180\, z^5\, s + 
            a\, h_{3}^2\, 135\, z^4\, (1 + 4\, s + 6\, s^2) + 
            f_{5}^2\, h_{3}\, 375\, s^2\, z^2\, (11 + 22\, z + 3\, z^2) + 
            a\, g_{5}\, h_{3}\, 90\, z^3\, (
                5\, z^3 + 30\, z^2\, s + 8\, s^2\, (2 + 5\, z)) + 
            b\, f_{6}\, h_{3}\, 720\, z^2\, (
                2 + 2\, s\, (1 + z^2) + s^2\, (7 + 3\, z^2)) + 
            a^2\, b\, h_{3}\, 432\, z^3\, (
                21\, z^2 + 21\, 5\, s + 5\, s^2\, (1 + 6\, z)))$,

$C[15, z_, s_] = \, 
      z/5\, (b\, f_{5}\, n_{3}\, (3000\, z^3 + 
                  4200\, z^3\, (1 + z) s) + n_{3}^2\, 45\, z^4\, s + 
  b\, g_{5}\, h_{3}\, \, 96\, z^3\, (7\, z^2 + 7\, 5\, s + 40\, s^2) + 
           a\, f_{6}\, n_{3}\, 540\, z^3\, ( z^2 + z\, 4\, s + 12\, s^2) + 
            b\, g_{5}^2\, 80\, z^2\, ( 
                14\, z^3 + 70\, z^2\, s + 75\, s^2\, (1 + 2\, z)) + 
            f_{5}\, f_{6}\, h_{3}\, 1800\, z^2\, (
                z^2 + 4\, z^2\, s + 6\, s^2\, (1 + 2\, z)) + 
            f_{6}^3\, 4320\, s^2\, (1 + 2\, z + 3\, z^2 + 4\, z^3) + 
            f_{5}\, f_{6}\, g_{5}\, 360\, z\, (
     13\, z^4 + 65\, z^3\, s +  50\, s^2\, (1 + 2\, z + 3\, z^2)) + 
            a_{24}\, h_{4}\, (80\, z^5 + 400\, z^5\, s) + 
            a_{15}\, h_{3}\, ( 105\, z^4 + 15\, s\, z^4\, (23 + 5\, z)) + 
 a_{05}\, h_{4}\, 100\, z^4\, s^2 + a\, h_{3}\, h_{4}\, 1800\, z^4\, s^2  + 
            f_{5}^2\, h_{4}\, 500\, z^2\, (
                6 + 6\, s\, (1 + z) + 5\, z^2\, s^2) + 
            a\, g_{5}\, h_{4}\, 24\, z^3\, (28\, z^2 + 28\, 5\, s\, z + 
                  25\, s^2\, (12 + 5\, z)) + 
            b\, f_{6}\, h_{4}\, 960\, z^2\, (
                2\, z^4 + 12\, s + 3\, s^2\, z\, (4 + z)) + 
            a^3\, n_{3}\, 3240\, z^4\, ( z + 4\, s + 8\, s^2) + 
  a^2\, f_{5}\, h_{3}\, 10800\, z^3\, (z^3  +  3\, (1 + s)\, s\, (1 + z)) + 
   a^2\, f_{6}^2\, 25920\, s^2\, z\, (3 + 6\, z + 9\, z^2 + 2\, z^3) + 
        a^2\, f_{5}\, g_{5}\, 720\, z^2\, (39\, z^3 + 5\, 39\, s\, z^3 + 
                  5\, s^2\, (30 + 60\, z + 59\, z^2))  + 
            a^2\, b\, h_{4}\, 5760\, z^3\, (
                2 + 6\, s + s^2\, (4 + 13\, z)) + 
            a^4\, f_{6}\, 155520\, s^2\, z^2\, (3 + 6\, z + 4\, z^2) + 
            a^6\, 933120\, s^2\, z^3\, (1 + 2\, z))$,

$C[16, z_, s_] = \, 
      z^2\, (3\, a_{15}\, h_{4}\, z^4 + 
  a_{24}\, b\, f_{5}\, 1040\, z^3\, s + a\, a_{24}\, f_{6}\, 936\, z^3\, s + 
            a\, h_{4}^2\, 24\, z^3\, ( 3 + 15\, s + 8\, s^2) + 
            f_{5}\, f_{6}\, h_{4}\, (
    29\, 60\, z^2 + 60\, (3 + s)\, s\, z\, (13 + 9\, z + 8\, z^2)) + 
            f_{6}^2\, g_{5}\, 180\, (
    8\, (1 + s\, z) + s^2\, (5 + 2\, z + 7\, z^2 + 12\, z^3)) + 
            b\, g_{5}\, h_{4}\, 40\, z^2\, (
                9\, z^2 + 48\, s + 4\, s^2\, (1 + 14\, z)) + 
            a^2\, f_{5}\, h_{4}\, 360\, z^2\, (
                29\, z^3 + s\, (52 + 52\, z + 21\, z^2))\,  + 
            f_{5}\, g_{5}^2\, 125\, z\, (
                3 + 6\, s + 2\, s^2\, (2 + 7\, z + 12\, z^2)) + 
            a^2\, f_{6}\, g_{5}\, 2160\, z\, (
                8 + 3\, s + s^2\, (2 + 7\, z + 12\, z^2)) + 
            a^4\, g_{5}\, 6480\, z^2\, (
                8\, z + s\, (1 + 5\, z) + 12\, s^2\, (1 + z))  + 
            a_{24}\, n_{3}\, ( 30\, z^4 + 36\, z^4\, (1 + z)\, s) + 
            f_{5}^2\, n_{3}\, 75\, z^2\, (
                6 + 18\, s\, z^2 + s^2\, (7 + 14\, z + 3\, z^2)) + 
            a\, g_{5}\, n_{3}\, 90\, z^3\, (
                1 + s\, (3 + z)\, (1 + 3\, s)) + 
            f_{5}\, h_{3}^2\, 45\, z^3\, (
                3 + 12\, s\, z + 2\, s^2\, (5 + 4\, z)) + 
            a\, h_{3}\, n_{3}\, 54\, z^4\, ( 1 + 5\, s + 7\, s^2) + 
            b\, h_{3}\, h_{4}\, 24\, z^3\, (
                9\, z + 48\, s + 4\, s^2\, (1 + 4\, z)) + 
            b\, f_{6}\, n_{3}\, 144\, z^2\, (
                6 + 18\, s\, z^2 + s^2\, (7 + 14\, z + 3\, z^2))  + 
            f_{6}^2\, h_{3}\, 108\, z\, (1 + z)\, (
                4\, z^4 + 13\, s\, (1 + z^2)) + 
            f_{5}\, g_{5}\, h_{3}\, 150\, z^2\, (
                3\, z^3 + 9\, s\, (1 + z) + 4\, s^2\, (1 + z)^2) + 
            a^2\, b\, n_{3}\, 864\, z^3\, (1 + z)\, (3\, z^2 + 13\, s) + 
            a^2\, f_{6}\, h_{3}\, 1296\, z^2\, (
                8\, z^4 + s\, (13 + 13\, z + 13\, z^2 + 4\, z^3)) + 
            a^4\, h_{3}\, 3888\, z^3\, (1 + z)\, ( 4\, z^2 + 13\, s))$,

$C[17, z_, s_] = \, 
      z^3\, (2100\, f_{6}\, g_{5}^2 + 12600\, a^2\, g_{5}^2\, z + 
    2520\, f_{6}\, g_{5}\, h_{3}\, z + 15120\, a^2\, g_{5}\, h_{3}\, z^2 + 
            756\, f_{6}\, h_{3}^2\, z^2 + 4536\, a^2\, h_{3}^2\, z^3 + 
            a^2\, a_{24}\, b\, ( 3456\, z^3 + 4608\, z^3\, s\, (1 + z)) + 
         a_{24}\, b\, f_{6}\, (576\, z^2 + 768\, z^2\, s\, (1 + z + z^2)) + 
            a\, a_{24}\, g_{5}\, (360\, z^3 + 480\, z^3\, s\, (1 + z)) + 
            a_{24}\, f_{5}^2\, (300\, z^2 + 400\, z^2\, s\, (1 + z + z^2)) + 
            a\, a_{24}\, h_{3}\, (216\, z^4 + 288\, z^4\, s) + 
            f_{5}\, h_{3}\, h_{4}\, (1680\, z^2\, s\, (1 + z) + 960\, z^5) + 
            b\, h_{4}^2\, 128\, z^2\, (z^3 + 7\, s + 7\, s^2\, z) + 
            f_{6}^2\, h_{4}\, (
                2016\, s\, (1 + z + z^2 + z^3) + 1152\, z^5) + 
            f_{5}\, g_{5}\, h_{4}\, (
                2800\, z\, s\, (1 + z + z^2) + 1600\, z^5) + 
            a^2\, f_{6}\, h_{4}\, (
                24192\, z\, s\, (1 + z + z^2) + 13824\, z^5) + 
            a^4\, h_{4}\, ( 72576\, z^2\, s\, (1 + z) + 41472\, z^5) + 
            a^3\, a_{15}\, 432\, s\, z^4\, (7 + 6\, z) + 
            a_{15}\, b\, f_{5}\, (
              80\, z^3\, s\, (1 + z) + 480\, z^3\, s\, (1 + z + z^2)) + 
            a\, a_{15}\, f_{6}\, (
                72\, z^3\, s\, (1 + z) + 
      432\, z^3\, s\, (1 + z + z^2)) + a_{15}\, n_{3}\, z^4\, s^2 + 
            b\, h_{3}\, n_{3}\, 24\, z^3\, (
                24\, z^3 + s\, (42 + 25\, z + 17\, z^2)) + 
            a\, h_{4}\, n_{3}\, 24\, z^3\, (
                6\, z^4 + s\, (7 + 3\, z)\, (1 + 5\, s + 2\, z^2)) + 
            b\, g_{5}\, n_{3}\, 40\, z^2\, (
                24\, z^4 + s\, (42 + 42\, z + 25\, z^2 + 17\, z^3)) + 
            f_{5}\, f_{6}\, n_{3}\, 60\, z\, (
   24\, z^5 + s\, (42 + 42\, z + 42\, z^2 + 25\, z^3 + 17\, z^4))  + 
            a^2\, f_{5}\, n_{3}\, 360\, z^2\, (
   24\, z^4 + s\, (42 + 42\, z + 25\, z^2 + 17\, z^3)))$,

$C[18, z_, s_] = \, 
      z^4\, (625\, g_{5}^3 + 3600\, f_{6}\, g_{5}\, h_{4} + 
       1125\, g_{5}^2\, h_{3}\, z + 21600\, a^2\, g_{5}\, h_{4}\, z + 
  2160\, f_{6}\, h_{3}\, h_{4}\, z + 1200\, f_{5}\, h_{4}^2\, z +
  675\, g_{5}\, h_{3}^2\, z^2 + 
            12960\, a^2\, h_{3}\, h_{4}\, z^2 + 135\, h_{3}^3\, z^3 + 
            a^2\, a_{24}\, f_{5}\, (
  4680\, z^2 + 1080\, z^2\, s\, (1 + z) + 5040\, z^2\, s\, (1 + z + z^2)) + 
            a^2\, a_{15}\, b\, (
                2880\, z^3\, s + 1440\, z^3\, s\, (1 + z)) + 
       a_{24}\, f_{5}\, f_{6}\, (780\, z + 180\, z\, s\, (1 + z + z^2) + 
                  840\, z\, s\, (1 + z + z^2 + z^3)) + 
            a_{24}\, b\, g_{5}\, (
                520\, z^2 + 120\, z^2\, s\, (1 + z) + 
                  560\, z^2\, s\, (1 + z + z^2)) + 
            a_{24}\, b\, h_{3}\, (
                312\, z^3 + 72\, z^3\, s + 336\, z^3\, s\, (1 + z)) + 
            a\, a_{24}\, h_{4}\, (312\, z^3 + 72\, z^3\, s + 
                  336\, z^3\, s\, (1 + z)) + 
            a_{15}\, b\, f_{6}\, (
      480\, z^2\, s\, (1 + z) + 240\, z^2\, s\, (1 + z + z^2)) + 
      a\, a_{15}\, g_{5}\, ( 300\, z^3\, s + 150\, z^3\, s\, (1 + z)) + 
            a_{15}\, f_{5}^2\, (
                250\, z^2\, s\, (1 + z) + 
                  125\, z^2\, s\, (1 + z + z^2)) + 
    a\, a_{15}\, h_{3}\, 90\, z^4\, s +
    a_{15}\, a_{24}\, (27\, z^4 + 2\, z^5) + 
            a\, n_{3}^2\, 27\, z^3\, (1 + 2\, s)\, (1 + 4\, s) + 
       f_{5}\, h_{3}\, n_{3}\, 90\, z^2\, (1 + z)\, (15\, s + 4\, z^2) + 
         b\, h_{4}\, n_{3}\, 96\, z^2\, (1 + z)\, (15\, s + 4\, z^2) + 
            f_{6}^2\, n_{3}\, 108\, (1 + z)\, (
                15\, s\, (1 + z^2) + 4\, z^4) + 
            f_{5}\, g_{5}\, n_{3}\, 150\, z\, (
                8\, z^4 + s\, (15 + 15\, z + 15\, z^2 + 4\, z^3)) + 
  a^2\, f_{6}\, n_{3}\, 1296\, z\, (4 + 11\, s\, (1 + z + z^2) + 4\, z^4) + 
            a^4\, n_{3}\, 3888\, z^2\, (1 + z)\, ( 15\, s + 4\, z^2) + 
            a_{05}\, a_{15}\, 5\, z^4\, (2 + z)\, (2 + 3  s))$,

$C[19, z_, s_] = \, 
      z^5\, (1600\, g_{5}^2\, h_{4} + 1536\, f_{6}\, h_{4}^2 + 
       2880\, f_{6}\, g_{5}\, n_{3} + 1920\, g_{5}\, h_{3}\, h_{4}\, z +
    9216\, a^2\, h_{4}^2\, z + 
    17280\, a^2\, g_{5}\, n_{3}\, z + 1728\, f_{6}\, h_{3}\, n_{3}\, z + 
       1920\, f_{5}\, h_{4}\, n_{3}\, z + 576\, h_{3}^2\, h_{4}\, z^2 + 
       10368\, a^2\, h_{3}\, n_{3}\, z^2 + 576\, b\, n_{3}^2\, z^2 + 
          a^4\, a_{24}\, 20736\, ( z^2 + z^2\, s\, (1 + z)) + 
          a^2\, a_{24}\, f_{6}\, 6912\, ( z + z\, s\, (1 + z + z^2)) + 
          a_{24}\, f_{5}\, g_{5}\, 800\, z\, ( 1 + s\, (1 + z + z^2)) + 
          a_{24}\, f_{6}^2\, 576\, ( 1 + s\, (1 + z + z^2 + z^3)) + 
          a_{24}\, b\, h_{4}\, 512\, z^2\, (1 + s\, (1 + z)) + 
          a_{24}\, f_{5}\, h_{3}\, 480\, z^2\, (1 + s\, (1 + z)) + 
          a\, a_{24}\, n_{3}\, 288\, z^3\, (1 + s) + 
          a^2\, a_{15}\, f_{5}\, (
 2520\, z^2 + 360\, z^2\, s\, (1 + z) + 2880\, z^2\, s (1 + z + z^2)) + 
            a_{15}\, f_{5}\, f_{6}\, (
 420\, z + 60\, z\, s\, (1 + z + z^2) + 480\, z\, s\, (1 + z + z^2 + z^3)) + 
            a_{15}\, b\, g_{5}\, (
 280\, z^2 + 40\, z^2\, s\, (1 + z) + 320\, z^2\, s\, (1 + z + z^2)) + 
 a_{15}\, b\, h_{3}\, (168\, z^3 + 24\, z^3\, s + 192\, z^3\, s\, (1 + z))  + 
 a\, a_{15}\, h_{4}\, (168\, z^3 + 24\, z^3\, s + 192\, z^3\, s\, (1 + z)))$,

$C[20, z_, s_] = \, 
      z^6\, (2040\, a_{24}\, f_{6}\, g_{5} + 
            1360\, g_{5}\, h_{4}^2 + 1275\, g_{5}^2\, n_{3} + 
 2448 \, f_{6}\, h_{4}\, n_{3} + 12240\, a^2\, a_{24}\, g_{5}\, z + 
 1224 \, a_{24}\, f_{6}\, h_{3}\, z + 1360\, a_{24}\, f_{5}\, h_{4}\, z + 
            816\, h_{3}\, h_{4}^2\, z + 1530\, g_{5}\, h_{3}\, n_{3}\, z + 
            14688\, a^2\, h_{4}\, n_{3}\, z + 765\, f_{5}\, n_{3}^2\, z + 
            7344\, a^2\, a_{24}\, h_{3}\, z^2 + 
            816\, a_{24}\, b\, n_{3}\, z^2 + 459\, h_{3}^2\, n_{3}\, z^2 + 
            204\, a\, a_{24}^2\, z^3 +  a^4\, a_{15}\, (
 10368\, z^2 + 6480\, z^2\, s\, (1 + z) + 5184\, z^2\, s\, (1 + z + z^2))  + 
            a^2\, a_{15}\, f_{6}\, (
 3456\, z + 2160\, z\, s\, (1 + z + z^2) +
 1728\, z\, s\, (1 + z + z^2 + z^3))  + 
            a_{15}\, f_{5}\, g_{5}\, (
                400\, z + 250\, z\, s\, (1 + z + z^2) + 
                  200\, z\, s\, (1 + z + z^2 + z^3)) + 
            a_{15}\, f_{6}^2\, (
                288 + 180\, s\, (1 + z + z^2 + z^3) + 
                  144\, s\, (1 + z + z^2 + z^3 + z^4))  + 
            a_{15}\, b\, h_{4}\, (
                256\, z^2 + 160\, z^2\, s\, (1 + z) + 
                  128\, z^2\, s\, (1 + z + z^2)) + 
            a_{15}\, f_{5}\, h_{3}\, (
                240\, z^2 + 150\, z^2\, s\, (1 + z) + 
                  120\, z^2\, s\, (1 + z + z^2)) + 
            a\, a_{15}\, n_{3}\, (
      144\, z^3 + 90\, z^3\, s + 72\, z^3\, s\, (1 + z) ))$,

$C[21, z_, s_] = \,
      1080\, a_{15}\, f_{6}\, g_{5}\, z^7 + 900\, a_{24}\, g_{5}^2\, z^7 + 
  1728\, a_{24}\, f_{6}\, h_{4}\, z^7 + 384\, h_{4}^3\, z^7 +
  2160\, g_{5}\, h_{4}\, n_{3}\, z^7 + 
    972\, f_{6}\, n_{3}^2\, z^7 + 6480\, a^2\, a_{15}\, g_{5}\, z^8 + 
 648\, a_{15}\, f_{6}\, h_{3}\, z^8 + 1080\, a_{24}\, g_{5}\, h_{3}\, z^8 + 
 10368\, a^2\, a_{24}\, h_{4}\, z^8 + 720\, a_{15}\, f_{5}\, h_{4}\, z^8 + 
 1080\, a_{24}\, f_{5}\, n_{3}\, z^8 + 1296\, h_{3}\, h_{4}\, n_{3}\, z^8 + 
        5832\, a^2\, n_{3}^2\, z^8 + 288\, a_{24}^2\, b\, z^9 + 
      3888\, a^2\, a_{15}\, h_{3}\, z^9 + 324\, a_{24}\, h_{3}^2\, z^9 + 
     432\, a_{15}\, b\, n_{3}\, z^9 + 216\, a\, a_{15}\, a_{24}\, z^{10}$,

$C[22, z_, s_] = \, 
      475\, a_{15}\, g_{5}^2\, z^8 + 912\, a_{15}\, f_{6}\, h_{4}\, z^8 + 
 1520\, a_{24}\, g_{5}\, h_{4}\, z^8 + 1368\, a_{24}\, f_{6}\, n_{3}\, z^8 + 
  912\, h_{4}^2\, n_{3}\, z^8 + 855\, g_{5}\, n_{3}^2\, z^8 +
 380\, a_{24}^2\, f_{5}\, z^9 + 
 570\, a_{15}\, g_{5}\, h_{3}\, z^9 + 5472\, a^2\, a_{15}\, h_{4}\, z^9 + 
 912\, a_{24}\, h_{3}\, h_{4}\, z^9 + 8208\, a^2\, a_{24}\, n_{3}\, z^9 + 
       570\, a_{15}\, f_{5}\, n_{3}\, z^9 + 513\, h_{3}\, n_{3}^2\, z^9 + 
   304\, a_{15}\, a_{24}\, b\, z^{10} + 171\, a_{15}\, h_{3}^2\, z^{10} + 
        57\, a\, a_{15}^2\, z^{11}$,

$C[23, z_, s_] = \, 
      480\, a_{24}^2\, f_{6}\, z^9 + 800\, a_{15}\, g_{5}\, h_{4}\, z^9 + 
        640\, a_{24}\, h_{4}^2\, z^9 + 720\, a_{15}\, f_{6}\, n_{3}\, z^9 + 
        1200\, a_{24}\, g_{5}\, n_{3}\, z^9 + 720\, h_{4}\, n_{3}^2\, z^9 + 
   2880\, a^2\, a_{24}^2\, z^{10} + 400\, a_{15}\, a_{24}\, f_{5}\, z^{10} + 
 480\, a_{15}\, h_{3}\, h_{4}\, z^{10} +
 4320\, a^2\, a_{15}\, n_{3}\, z^{10} + 
    720\, a_{24}\, h_{3}\, n_{3}\, z^{10} + 80\, a_{15}^2\, b\, z^{11}$,

$C[24, z_, s_] = \, 
  504\, a_{15}\, a_{24}\, f_{6}\, z^{10} + 420\, a_{24}^2\, g_{5}\, z^{10} + 
  336\, a_{15}\, h_{4}^2\, z^{10} + 630\, a_{15}\, g_{5}\, n_{3}\, z^{10} + 
      1008\, a_{24}\, h_{4}\, n_{3}\, z^{10} + 189\, n_{3}^3\, z^{10} + 
  3024\, a^2\, a_{15}\, a_{24}\, z^{11} + 105\, a_{15}^2\, f_{5}\, z^{11} + 
  252\, a_{24}^2\, h_{3}\, z^{11} + 378\, a_{15}\, h_{3}\, n_{3}\, z^{11}$,

$C[25, z_, s_] = \,
 132\, a_{15}^2\, f_{6}\, z^{11} + 440\, a_{15}\, a_{24}\, g_{5}\, z^{11} + 
 352\, a_{24}^2\, h_{4}\, z^{11} + 528\, a_{15}\, h_{4}\, n_{3}\, z^{11} + 
      396\, a_{24}\, n_{3}^2\, z^{11} + 792\, a^2\, a_{15}^2\, z^{12} + 
      264\, a_{15}\, a_{24}\, h_{3}\, z^{12}$,

$C[26, z_, s_] = 
 115\, a_{15}^2\, g_{5}\, z^{12} + 368\, a_{15}\, a_{24}\, h_{4}\, z^{12} + 
       276\, a_{24}^2\, n_{3}\, z^{12} + 207\, a_{15}\, n_{3}^2\, z^{12} + 
        69\, a_{15}^2\, h_{3}\, z^{13}$,

$C[27, z_, s_] = \, 
      64\, a_{24}^3\, z^{13} + 96\, a_{15}^2\, h_{4}\, z^{13} + 
        288\, a_{15}\, a_{24}\, n_{3}\, z^{13}$,

$C[28, z_, s_] = \, 
      100\, a_{15}\, a_{24}^2\, z^{14} + 75\, a_{15}^2\, n_{3}\, z^{14}$,

$C[29, z_, s_] = \, 52\, a_{15}^2\, a_{24}\, z^{15}$,

$C[30, z_, s_] = \, 9\, a_{15}^3\, z^{16}$,
\end{flushleft}

\end{document}